\RequirePackage{fix-cm}
\documentclass[twocolumn,epjc3]{svjour3}
\smartqed  
\RequirePackage{graphicx}
\RequirePackage{mathptmx}      

\usepackage{color,graphicx,epsfig}
\usepackage{ifpdf}
\usepackage{amsmath}
\usepackage{bm}
\usepackage{color}
\usepackage[english]{babel}
\usepackage{graphicx}%
\usepackage{amsfonts}%
\usepackage{amssymb}
\usepackage{braket}
\usepackage{hyperref}

\definecolor{nicered}{rgb}{0.7,0.1,0.1}
\definecolor{nicegreen}{rgb}{0.1,0.5,0.1}
\hypersetup{colorlinks,citecolor= nicegreen,linkcolor= nicered}

\newcommand{\beq}{\begin{equation}}
\newcommand{\eeq}{\end{equation}}
\newcommand{\bea}{\begin{eqnarray}}
\newcommand{\eea}{\end{eqnarray}}

\newcommand{\arXhref}[1]{\href{http://arxiv.org/abs/#1}{#1}}
\journalname{Eur. Phys. J. C}

\usepackage{xspace}
\newcommand{\FeynRules}{{\rmfamily\scshape FeynRules}\xspace}
\newcommand{\MadGraph}{{\rmfamily\scshape MadGraph5\_aMC@NLO}\xspace}

\begin{document}

\def\Zurich{Physik-Institut, Universit\"at Z\"urich, CH-8057 Z\"urich, Switzerland}
\def\Sarajevo{Faculty of Science, University of Sarajevo, Zmaja od Bosne 33-35, 71000 Sarajevo, Bosnia and Herzegovina}

\title{
High-$p_T$ dilepton tails and flavour physics
}

\author{Admir Greljo\thanksref{addr1,addr2} and David Marzocca\thanksref{addr1}}
\institute{\Zurich 
\label{addr1}
\and
\Sarajevo \label{addr2}}

\maketitle

\begin{abstract}
{
We investigate the impact of flavour-conserving, non-universal quark-lepton contact interactions on the dilepton invariant mass distribution in $p~p \to \ell^+ \ell^-$ processes at the LHC. After recasting the recent ATLAS search performed at 13~TeV with $36.1$~fb$^{-1}$ of data, we derive the best up-to-date limits on the full set of  36 relevant four-fermion operators, as well as estimate the sensitivity achievable at the HL-LHC.
We discuss how {these high-$p_T$ measurements can provide complementary information to} the low-$p_T$ rare meson decays. In particular, {we find that the} recent hints on lepton flavour universality violation in $b \to s \mu^+ \mu^-$ transitions are already in mild tension with the dimuon spectrum at high-$p_T$ if the flavour structure follows minimal flavour violation. Even if the mass scale of New Physics is well beyond the kinematical reach for on-shell production, the signal in the high-$p_T$ dilepton tail might still be observed, a fact that has been often overlooked in the present literature. In scenarios where new physics couples predominantly to third generation quarks, instead, the HL-LHC phase is necessary in order to provide valuable information.
}

\end{abstract}

\section{Introduction}
\label{sec:I}

Searches for new physics in flavour-changing neutral currents (FCNC) at low energies set strong limits on flavour-violating semi-leptonic four-fermion operators ($qq'\ell\ell$), {often} pushing the new physics mass scale $\Lambda$ beyond the kinematical reach of the LHC~\cite{Isidori:2010kg}. For example, if the recent hints for lepton flavour non-universality in $b \to s \ell^+ \ell^-$ transitions~\cite{Aaij:2014ora,Aaij:2013qta,Aaij:2015oid,BifaniTALK} are confirmed, the relevant dynamics might easily be outside the LHC range for on-shell production.

In this situation, an effective field theory (EFT) approach is applicable in the entire spectrum of momentum transfers in proton collisions at the LHC, including the most energetic processes.  Since the leading deviations from the SM scale like $\mathcal{O}(p^{2}/\Lambda^{2})$, where $p^2$ is a typical momentum exchange, less precise measurements at high-$p_T$ could offer similar (or even better) sensitivity to new physics with respect to
high-precision measurements at low energies. Indeed, opposite-sign same-flavour charged lepton production, $p~p \to \ell^+ \ell^-$ ($\ell=e, \mu$), sets competitive constraints on new physics when compared to some low-energy measurements~\cite{Cirigliano:2012ab,Gonzalez-Alonso:2016sip,deBlas:2013qqa} or electroweak precision tests performed at LEP~\cite{Farina:2016rws}.

At the same time, motivated new physics flavour structures can allow for large flavour-conserving but flavour non-universal interactions. In this work we study the impact of such contact interactions on the tails of dilepton invariant mass distribution in $p~p \to \ell^+ \ell^-$ and use the limits obtained in this way to derive bounds on class of models which aim to solve the recent $b \to s \ell\ell$ anomalies.
With a similar spirit, in Ref.~\cite{Faroughy:2016osc} it was shown that the LHC measurements of $p p \to \tau^+ \tau^-$ already set stringent constraints on models aimed at solving the charged-current $b \to c \tau \bar{\nu}_\tau$ anomalies.
The paper is organized as follows. In Sec.~\ref{sec:II} we present a general parameterisation of new physics effects in $p~p \to \ell^+ \ell^-$ and perform a recast of the recent ATLAS search at 13~TeV with 36.1~fb$^{-1}$ of data~\cite{ATLAS:2017wce} to derive present and future projected limits on flavour non-universal contact interactions for all quark flavours accessible in the initial protons. In Sec.~\ref{sec:flavour} we discuss the implications of these results on the rare FCNC $B$ meson decay anomalies. We conclude in Sec.~\ref{sec:conclusions}.

\section{New physics in the dilepton tails}
\label{sec:II}

\subsection{General considerations}

Let us start the discussion on new physics contributions to dilepton production via Drell-Yan by listing the gauge-invariant dimension-six operators which can contribute at tree-level to the process.
We opt to work in the Warsaw basis~\cite{Grzadkowski:2010es}. 
Neglecting chirality-flipping interactions (e.g. scalar or tensor currents, expected to be suppressed by the light fermion Yukawa couplings), dimension-six operators can contribute to $q~ \bar{q} \to \ell^+ \ell^-$ either by modifying the SM contributions due to the $Z$ exchange, or via local four-fermion interactions.
The former class of deviations can be probed with high precision by on-shell $Z$ production and decays at both LEP-1 and LHC (see e.g. Ref.~\cite{Efrati:2015eaa}). Also, such effects are not enhanced at high energies, scaling like $\sim v^2 / \Lambda^2$.
Therefore we neglect them and focus on the four-fermion interactions which comprise of four classes depending on the chirality: $(\bar L L)(\bar L L)$, $(\bar R R)(\bar R R)$, $(\bar R R)(\bar L L)$, and $(\bar L L)(\bar R R)$.
In particular, the relevant set of operators is:
{\footnotesize
\bea
&& \mathcal L^{\rm SMEFT} \supset \nonumber \\
&& \frac{c^{(3)}_{Q_{ij} L_{kl}}}{\Lambda^2} (\bar Q_i \gamma_\mu \sigma^a Q_j) (\bar L_k \gamma^\mu \sigma_a L_l)+\frac{c^{(1)}_{Q_{ij} L_{kl}}}{\Lambda^2} (\bar Q_i \gamma_\mu Q_j) (\bar L_k \gamma^\mu L_l) + \nonumber\\
	&& \frac{c_{u_{ij} e_{kl}}}{\Lambda^2} (\bar u_i \gamma_\mu u_j) (\bar e_k \gamma^\mu e_l) 
	+\frac{c_{d_{ij} L_{kl}}}{\Lambda^2} (\bar d_i \gamma_\mu d_j) (\bar e_k \gamma^\mu e_l) + \nonumber\\
	&& \frac{c_{u_{ij} L_{kl}}}{\Lambda^2} (\bar u_i \gamma_\mu u_j) (\bar L_k \gamma^\mu L_l) 
	+\frac{c_{d_{ij} L_{kl}}}{\Lambda^2} (\bar d_i \gamma_\mu d_j) (\bar L_k \gamma^\mu L_l) + \nonumber\\
	&& \frac{c_{Q_{ij} e_{kl}}}{\Lambda^2} (\bar Q_i \gamma_\mu Q_j) (\bar e_k \gamma^\mu e_l)
	\label{eq:opSMEFT}
\eea}
\!\!\!where {$i,j,k,l$ are flavour indices,} $Q_i = ({V_{ji}^*} u^j_{L},d^i_L)^T$ and $L_i = (\nu^i_L,\ell^i_L)^T$ are the SM left-handed quark and lepton weak doublets, while $d_i$, $u_i$, $e_i$ are the right-handed singlets. $V$ is the CKM flavour mixing matrix and $\sigma^a$ are the Pauli matrices acting on $SU(2)_L$ space.

An equivalent classification of the possible contact interactions can be obtained by studying directly the $q  ~\bar q \to \ell^- \ell^+$ scattering amplitude:
\beq
	\mathcal{A}(q^i_{p_1}\bar{q}^j_{p_2} \to \ell^{-}_{p'_1} \ell^{+}_{p'_2}) = i \sum_{q_L,q_R}\sum_{\ell_L, \ell_R} (\bar q^i \gamma^\mu q^j) ~ (\bar \ell \gamma_\mu \ell) ~ F_{q \ell}(p^2)~, \label{eq:1}
\eeq
where $p\equiv p_1+p_2=p'_1+p'_2$, {and the form factor $F_{q \ell}(p^2)$ can be} expanded around the propagating physical poles (photon and $Z$ boson), leading to
\beq
	F_{q \ell}(p^2) = \delta^{ij}\frac{e^2 Q_q Q_\ell}{p^2} + \delta^{ij}\frac{ g_Z^q g_Z^\ell}{p^2-m_Z^2 + i m_Z \Gamma_Z}+ {\frac{\epsilon^{q\ell}_{ij}}{v^2}}~.
\eeq
Here, $Q_{q(\ell)}$ is the quark (lepton) electric charge, while $g_Z^{q(\ell)}$ is the corresponding coupling to $Z$ boson: in the SM\\ {\footnotesize $g_Z^{f} = \frac{2 m_Z}{v} (T^3_f - Q_f \sin^2 \theta_W)$}. 
The contact terms $\epsilon_{ij}^{q\ell}$ are related to the EFT coefficients in Eq.~\eqref{eq:opSMEFT} by simple relations $\epsilon_x = \frac{v^2}{\Lambda^2} c_x$, with $v \simeq 246$~GeV. The only constraint on the contact terms imposed by $SU(2)_L$ invariance are $\epsilon^{d_L e_R^k}_{ij} = \epsilon^{u_L e_R^k}_{ij} = c_{Q_{ij} e_{kk}} v^2 / \Lambda^2$.

\begin{figure}[t]
\centering
\includegraphics[width=1.0\hsize]{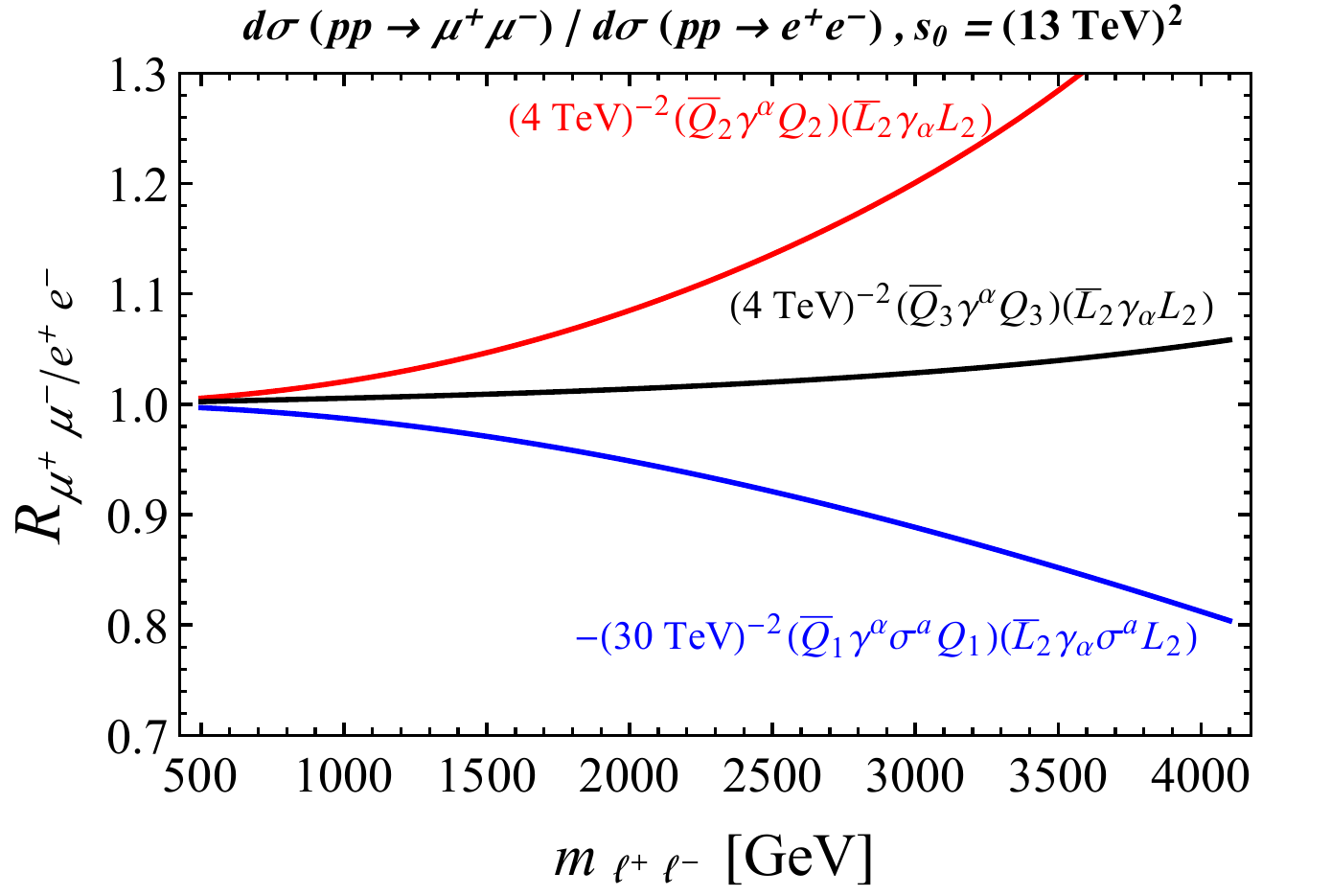}\\
\caption{$R_{\mu^+ \mu^-/e^+ e^-}$ as a function of the dilepton invariant mass $ m_{\ell^+ \ell^-}$ for three new physics benchmark points. See text for details. }\label{fig:R-plot}
\end{figure}

The dilepton invariant mass spectrum can be written as (see \ref{app:sigma}),
\beq
	\frac{d \sigma}{d \tau} = \left( \frac{d \sigma}{d \tau} \right )_{\rm{SM}} \times \frac{\sum_{q,\ell} \mathcal{L}_{q \bar q}(\tau,\mu_F) |F_{q \ell}(\tau s_0)|^2}{\sum_{q,\ell} \mathcal{L}_{q \bar q}(\tau,\mu_F) |F_{q \ell}^{\rm{SM}}(\tau s_0)|^2}~,
\eeq
where  $\tau\equiv m_{\ell^+ \ell^-}^2/s_0$ and $\sqrt{s_0}$ is the proton-proton center of mass energy. The sum is over the left- and right-handed quarks and leptons as well as the quark flavours accessible in the proton.
Note that, {since we are interested in the high-energy tails (away from the $Z$ pole),} the universal higher-order radiative QCD corrections factorize (to a large extent).
Therefore, consistently including those corrections in the SM prediction is enough to achieve good theoretical accuracy.
It is still useful to define the differential LFU ratio, 
\beq\begin{split}
	&R_{\mu^+ \mu^-/e^+ e^-} (m_{\ell\ell}) \equiv \frac{ d \sigma_{\mu\mu}}{ d m_{\ell\ell} } / \frac{ d \sigma_{ee} }{ d m_{\ell\ell} } = \\
	&\quad = \frac{\sum_{q,\mu} \mathcal{L}_{q \bar q}(m_{\ell \ell}^2/s_0,\mu_F) |F_{q \mu}(m_{\ell\ell}^2)|^2}{\sum_{q,e} \mathcal{L}_{q \bar q}(m_{\ell\ell}^2/s_0,\mu_F) |F_{q e}(m_{\ell\ell}^2)|^2}~,
\end{split}\eeq
which is a both theoretically and experimentally cleaner observable. As an illustration, we show in Fig.~\ref{fig:R-plot} the predictions for $R_{\mu^+ \mu^-/e^+ e^-}$ at $\sqrt{s_0} =13$~TeV, assuming new physics in three benchmark operators. The parton luminosities used to derive these predictions are discussed in the next chapter.  

A goal of this work is to connect the high-$p_T$ dilepton tails measurements with the recent experimental hints on lepton flavour universality violation in rare semileptonic $B$ meson decays.
The pattern of observed deviations points towards new physics contributions in left-handed quark currents involving muons, as discussed in the next section in more details.
For this reason, when discussing the connection to flavour in Section~\ref{sec:flavour}, we limit our attention to the $(\bar L L)(\bar L L)$ operators with muons given in the first line of Eq.~\eqref{eq:opSMEFT}.
For this purpose, it is useful to rearrange the terms relevant to $p ~p \to \mu^+ \mu^-$ as:\footnote{{The down and up couplings are given by two orthogonal combinations of the triplet and singlet operators in the first line of Eq.~\eqref{eq:opSMEFT}: ${\bf C}^{D(U)\mu}_{ij} = v^2 / \Lambda^2 (c^{(1)}_{Q_{ij} L_{22}} \pm c^{(3)}_{Q_{ij} L_{22}} )$.}}
{\small
\beq
	\mathcal L^{\rm eff} \supset \frac{{\bf C}^{U\mu}_{ij}}{v^2} (\bar u^i_L \gamma_\mu u^j_L) (\bar \mu_L \gamma^\mu \mu_L)+ \frac{{\bf C}^{D\mu}_{ij}}{v^2} (\bar d^i_L \gamma_\mu d^j_L) (\bar \mu_L \gamma^\mu \mu_L)~,
\label{eq:EFTpp2mu}
\eeq}
\!\!The ${\bf C}^{U\mu}$ and ${\bf C}^{D\mu}$ matrices carry the flavour structure of the operators. Since the top quark does not appear in the process under study we can neglect the corresponding terms. Regarding the off-diagonal elements, we keep only the $b-s$ one since it is where the flavour anomalies appear, while we set the others to zero. In summary:
\beq
	{\bf C}^{U\mu}_{ij} = \left( \begin{array}{ccc}
	C_{u\mu} & 0 & 0 \\
	0 & C_{c\mu} & 0 \\
	0 & 0 & C_{t\mu} 
	\end{array}\right), ~~
	{\bf C}^{D\mu}_{ij} = \left( \begin{array}{ccc}
	C_{d\mu} & 0 & 0 \\
	0 & C_{s\mu} & C_{bs\mu}^* \\
	0 & C_{bs\mu} & C_{b\mu} 
	\end{array}\right).
	\label{eq:EFTpp2mu2}
\eeq

\subsection{Present limits and HL-LHC projections}
\label{sec:limits}

In this section we derive limits on the flavour non-universal quark-lepton contact interactions by looking in the tails of dilepton invariant mass distributions in $p~p \to \ell^+ \ell^-$ at the LHC. In our analysis, we closely follow the recent ATLAS search~\cite{ATLAS:2017wce} performed at 13~TeV with 36.1 fb$^{-1}$ of data. We digitise Figure~1 of Ref.~\cite{ATLAS:2017wce}, which shows the distribution of dielectron and dimuon reconstructed invariant masses after the final event selection.
We perform a profile likelihood fit to a binned histogram distribution adopting the method from Ref.~\cite{Cowan:2010js}. The number of signal events, as well as the expected signal events in the SM and background processes, are directly taken from the  Figure~1 of Ref.~\cite{ATLAS:2017wce}. The likelihood function ($L$) is constructed treating every bin as an independent Poisson variable, with the expected number of events,
{\small
\beq
	\Delta N^{\rm{bin}} = \Delta N^{\rm{bin}}_{\rm{SM}} \times \frac{\sum_{q,\ell} \int_{\tau_{\rm{min}}^{\rm{bin}}}^{\tau_{\rm{max}}^{\rm{bin}}} d\tau ~ \tau ~\mathcal{L}_{q \bar q}(\tau,\mu_F) ~|F_{q \ell}(\tau s_0)|^2}{\sum_{q,\ell} \int_{\tau_{\rm{min}}^{\rm{bin}}}^{\tau_{\rm{max}}^{\rm{bin}}} d\tau ~ \tau ~\mathcal{L}_{q \bar q}(\tau,\mu_F) ~|F_{q \ell}^{\rm{SM}}(\tau s_0)|^2}~,
\label{Eq:deltaN}
\eeq}
\!\!which is a function of the contact interactions.
The best fit point corresponds to the global minimum of $\chi^{2} \equiv - 2 \log L$, while $n\sigma$ C.L. regions are given as $\Delta\chi^{2}\equiv\chi^{2}-\chi^{2}_{\rm{min}}<\Delta_{n \sigma}$, where $\Delta_{n \sigma}$ are defined with the appropriate cumulative distribution functions.
In the numerical study, we use the NNLO$_{118}$ MMHT2014 parton distribution functions set~\cite{Harland-Lang:2014zoa}. We checked that {our results} have a very small dependence on the factorization scale variation.

\begin{figure}[t]
\centering
\includegraphics[width=0.9\hsize]{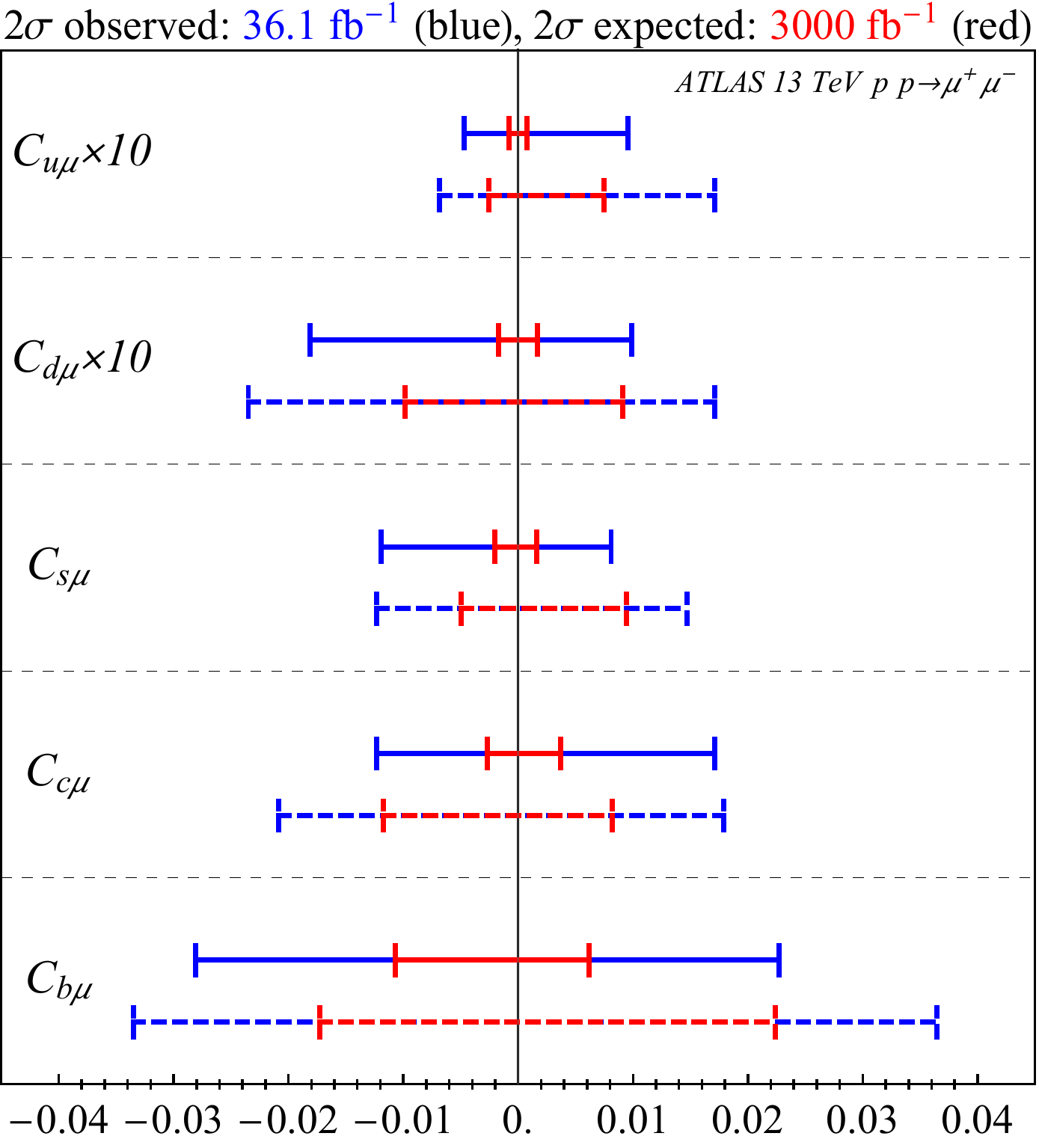}
\caption{In blue (red) we show the present (projected) $2\sigma$ limits on $C_{q\mu}$ (flavour conserving $(\bar L L)(\bar L L)$ operators) where $q = u,d,s,c$ and $b$, using 13~TeV ATLAS search in $p p \to \mu^+ \mu^-$ channel~\cite{ATLAS:2017wce}. Dashed lines show the limits when all other coefficients are marginalised, while the solid ones show the results of one-parameter fits.}\label{fig:FIT1d}
\end{figure}

Furthermore, we independently cross-check the results by implementing the subset of operators in Eqs.~(\ref{eq:EFTpp2mu},\ref{eq:EFTpp2mu2}) in a \FeynRules \cite{Alloul:2013bka} model, and generating $p p \to \mu^+ \mu^-$ events at 13 TeV with the same acceptance cuts as in the ATLAS search~\cite{ATLAS:2017wce} using \MadGraph~\cite{Alwall:2014hca}. We find good agreement between the fits performed in both ways.

In the SMEFT, neglecting flavour-violating interactions, there are 18 independent four-fermion operators for muons and 18 for electrons relevant to $p p \to \ell^+ \ell^-$ (see Eq.~\eqref{eq:opSMEFT}). In \ref{app:Limits} (Tab.~\ref{tab:Fit1}) we provide present and projected $2\sigma$ limits on all these coefficients, using the recent ATLAS search~\cite{ATLAS:2017wce}. While these limits are obtained in the scenario where only one operator is considered at a time, we checked that the $18\times18$ correlation matrix derived in the Gaussian approximation does not contain any large value (the only non-negligible correlations are among the triplet and singlet operators with the same flavour content, which is discussed in more details below). The absence of flat directions can be understood by the fact that operators with fermions of different flavour or chirality do not interfere with each other.

Focusing only on the $(\bar L L)(\bar L L)$ operators (in the notation of Eq.~\eqref{eq:EFTpp2mu}), the 2$\sigma$ limits, both from the present ATLAS search (blue) and projected for 3000 fb$^{-1}$ (red), are shown in Fig.~\ref{fig:FIT1d}. The solid lines show the 2$\sigma$ bounds when operators are taken one at a time, while the dashed ones show the limits when all the others are marginalised. The small difference between the two, especially with present accuracy, confirms what we commented above.

\section{Implications for $R(K)$ and $R(K^*)$}
\label{sec:flavour}

\subsection{Effective field theory discussion}

Recent measurements in rare semileptonic $b \to s$ transitions provide strong hints for a new physics contribution to $b s \mu \mu$ local interactions (see for example the recent analyses in Refs.~\cite{Capdevila:2017bsm,Altmannshofer:2017yso,Geng:2017svp}).
In particular, a good fit of the anomaly in the differential observable $P_5'$~\cite{DescotesGenon:2012zf}, together with the hints on LFU violation in $R_K$ and $R_{K^*}$~\cite{Hiller:2003js,Bobeth:2007dw,Bordone:2016gaq}, is obtained by considering a new physics contribution to the $C_{bs\mu}$ coefficient in Eqs.~(\ref{eq:EFTpp2mu},\ref{eq:EFTpp2mu2}). In terms of the SMEFT operators at the electroweak scale, this corresponds to a contribution to (at least) one of the two operators in the first row of Eq.~\eqref{eq:opSMEFT} (see for example~\cite{Celis:2017doq}). Moreover, the triplet operator could at the same time solve the anomalies in charged-currrent ($R_{D^{(*)}}$) , see e.g. Refs.~\cite{Bhattacharya:2014wla,Alonso:2015sja,Greljo:2015mma}.

Matching at the tree level this operator to the standard effective weak Hamiltonian describing $b \to s$ transitions, one finds
\beq
	\Delta C_9^{\mu} = - \Delta C_{10}^\mu = \frac{\pi}{\alpha V_{t b} V^*_{t s}} C_{bs\mu}~,
	\label{eq:matching}
\eeq
where $\alpha$ is the electromagnetic fine structure constant while $|V_{ts}| = (40.0 \pm 2.7) \times 10^{-3}$ and $|V_{tb}| = 1.009 \pm 0.031$ are CKM matrix elements~\cite{Olive:2016xmw}.

The recent combined fit of Ref.~\cite{Capdevila:2017bsm} reported the best fit value and $1\sigma$ preferred range
\beq
	\Delta C_9^{\mu} = - \Delta C_{10}^\mu = -0.61 \pm 0.12~.
	\label{eq:DC9}
\eeq
Using this result and Eq.~\eqref{eq:matching}, one can estimate the scale of the relevant new physics by defining $C_{bs\mu} = g_*^2  v^2 / \Lambda^2$, obtaining $\Lambda / g_* \approx 32^{+4}_{-3}$~TeV. Depending on the value of $g_*$, i.e. from the particular UV origin of the operator, the scale of new physics $\Lambda$ can be within or out of the reach of LHC direct searches. We show that even in the latter case, under some assumptions it can be possible to observe an effect in the dimuon high energy tail. When comparing low and high-energy measurements, the renormalisation group effects should in principle be taken into account. Since these effects are small in this case, we neglect it in what follows (see for example~\cite{Celis:2017doq}).

We concentrate on UV models in which new particles are above the scale of threshold production at the LHC, such that the EFT approach is applicable in the most energetic dilepton events. We stress however that even for models with light new physics these searches can be relevant.

Let us discuss the flavour structure of the ${\bf C}^{D(U)\mu}_{ij}$ matrices in Eqs.~(\ref{eq:EFTpp2mu},\ref{eq:EFTpp2mu2}).
New physics aligned only to the strange-bottom coupling $C_{bs\mu}$ will not be probed at the LHC, in fact the present (projected) 95\% CL limits from the 13 TeV ATLAS $pp \to \mu^+\mu^-$ analysis with 36~fb$^{-1}$ (3000~fb$^{-1}$) of luminosity are
\beq
	\left| \frac{\pi}{\alpha V_{t b} V^*_{t s}} C_{bs\mu} \right| < 100~ (39)~,
	\label{eq:Cbslimit}
\eeq
which should be compared with the value extracted from the global flavour fits in Eq.~\eqref{eq:DC9}.
Such a peculiar flavour structure is possible, but not very motivated from the model building point of view.

On the other hand, taking the $b \to s \mu^+ \mu^-$ flavour anomalies at face value provides a measurement of the $C_{bs\mu}$ coefficient (via Eq.~\eqref{eq:matching}).
In most flavour models flavour-violating couplings are related (by symmetry or dynamics) to flavour-diagonal one(s).
In this case we can use the LHC upper limit on $|C_{q\mu}|$ from the dimuon high-$p_T$ tail in order to set a lower bound on {$|\lambda^q_{bs}|$, defined as the ratio}
\beq
	\lambda^q_{bs} \equiv C_{bs\mu} / C_{q\mu}~.
\eeq
In the following we study such limits for several particularly interesting scenarios.

\begin{figure}[t]
\centering
\includegraphics[width=0.9\hsize]{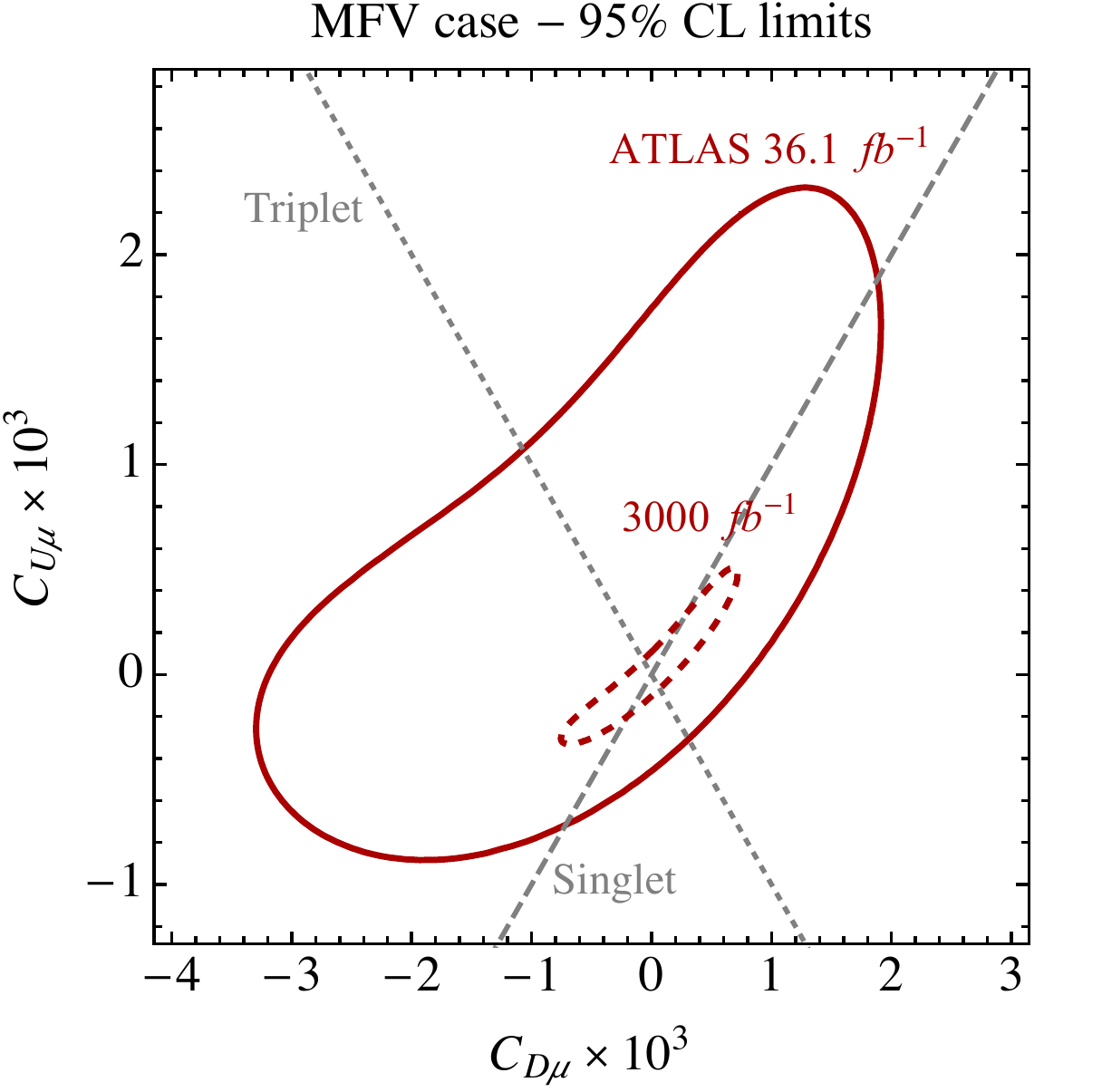}
\caption{Present and projected 95\% CL limits from $p p \to \mu^+ \mu^-$ in the MFV case defined by Eq.~\eqref{eq:MFVstructure}.}\label{fig:MFV}
\end{figure}

\begin{figure}[t]
\centering
\includegraphics[width=0.95\hsize]{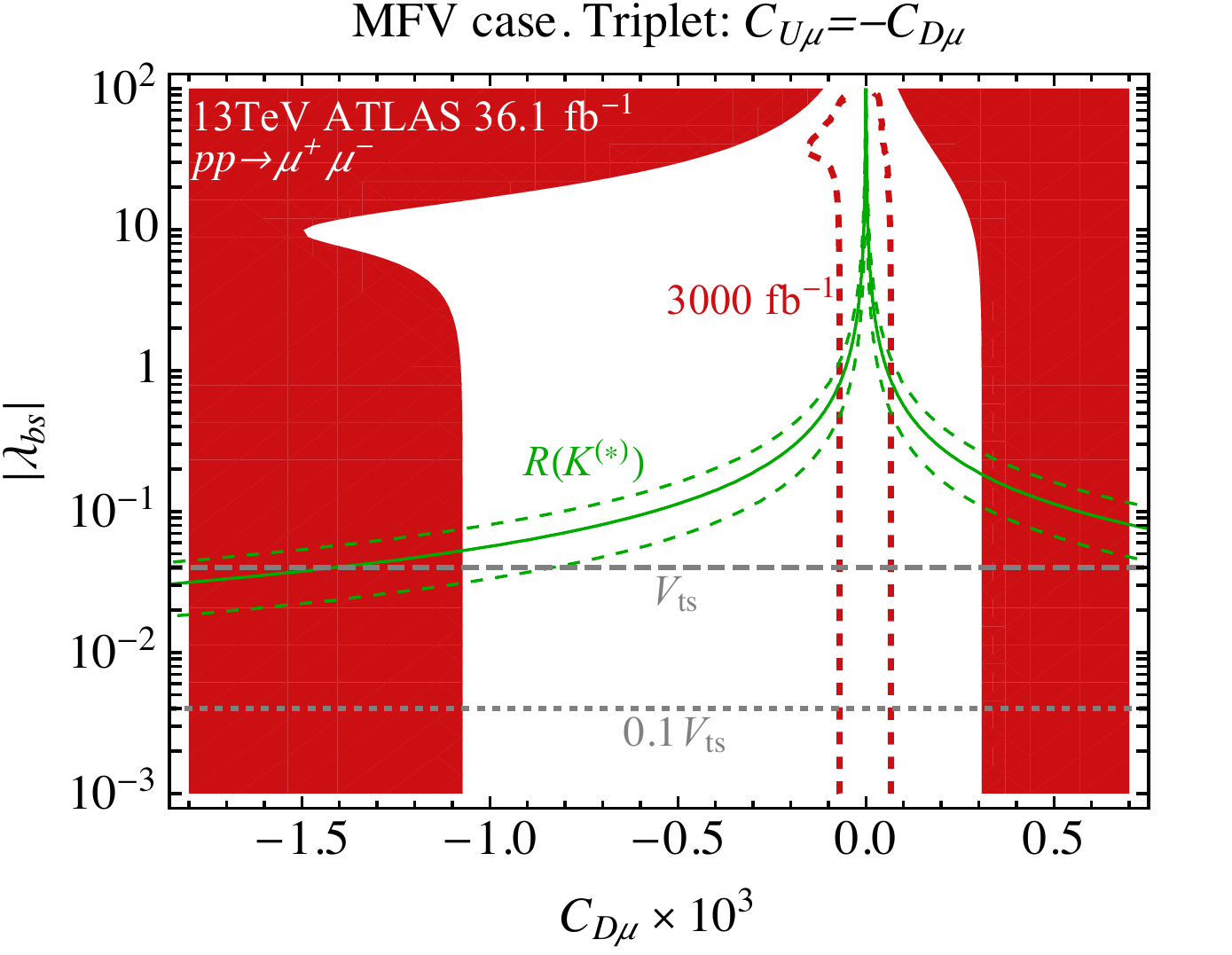}\\
\includegraphics[width=0.95\hsize]{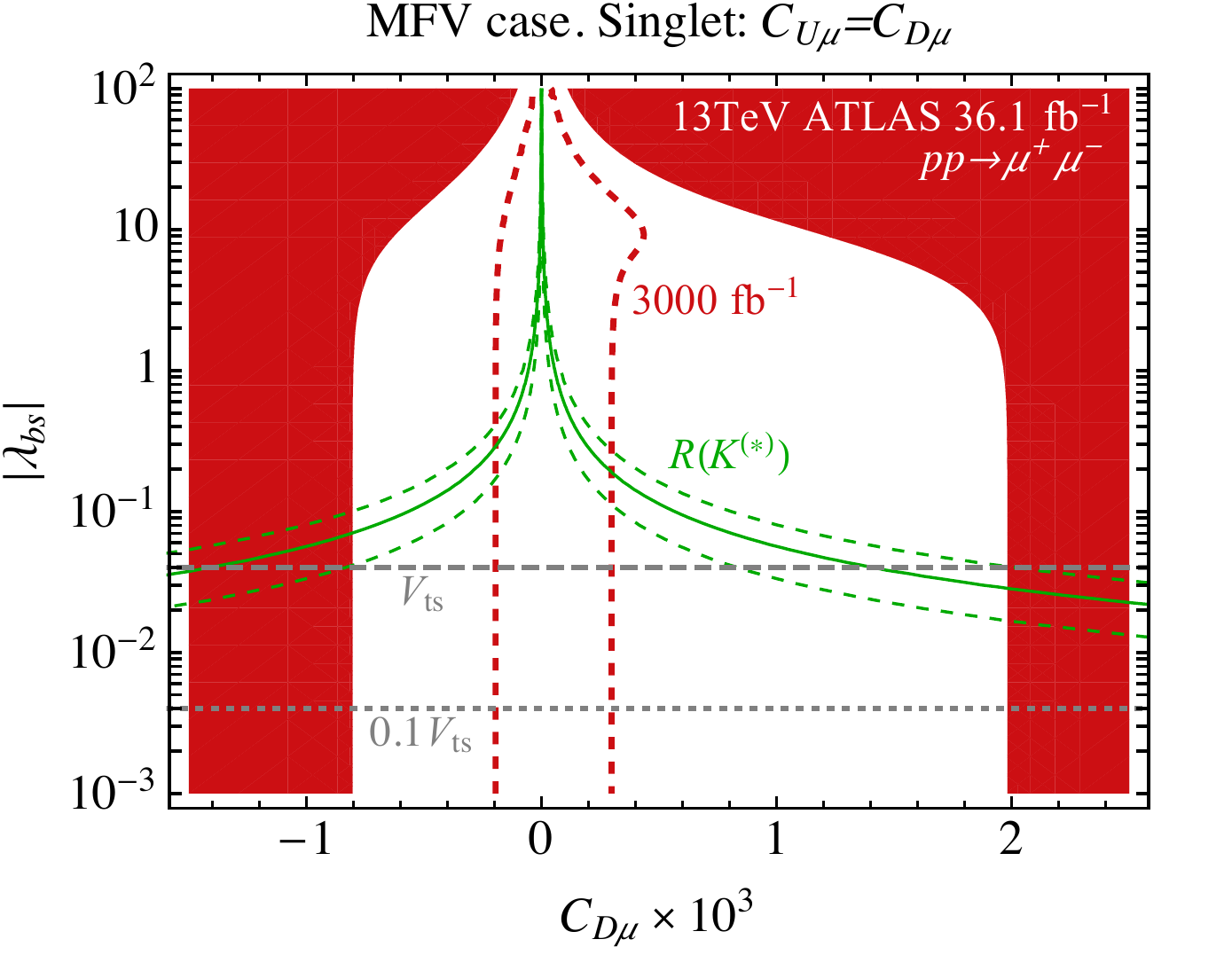}\\
\includegraphics[width=0.95\hsize]{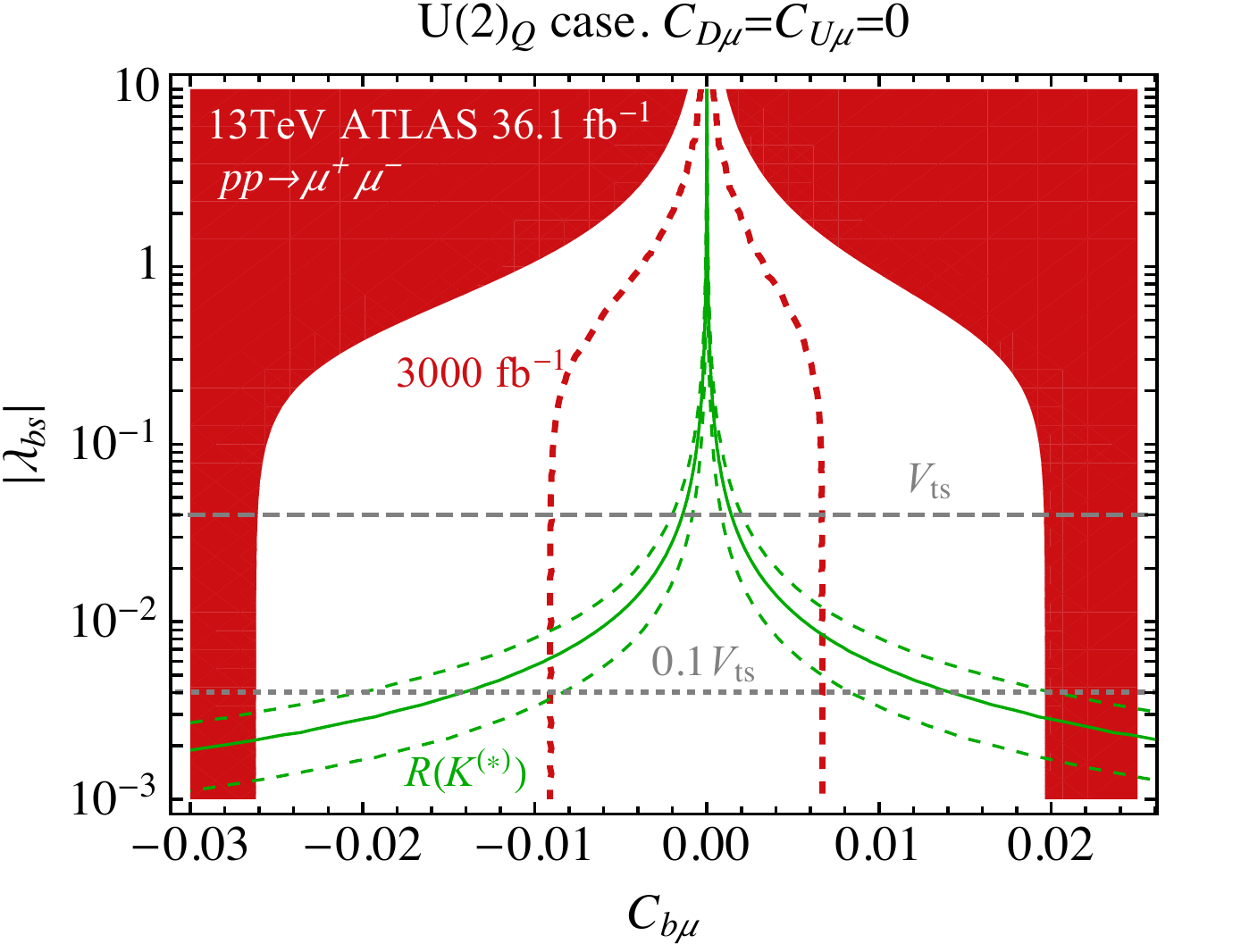}
\caption{We show the present (solid red) and projected (dashed red) 95\% CL limit from $p p \to \mu^+ \mu^-$ in the $C_{q\mu}$-$|\lambda_{bs}|$ plane. The solid (dashed) green line corresponds to the best fit (2$\sigma$ interval) from the fit of the flavour anomalies in Eq.~\eqref{eq:DC9}.}\label{fig:curtain}
\end{figure}

{\bf 1) Minimal flavour violation}\\
Under this assumption~\cite{DAmbrosio:2002vsn} the only source of flavour violation are the SM Yukawa matrices $Y_u \equiv V^\dagger {\rm diag}(y_u,y_c,y_t)$ and $Y_d \equiv {\rm diag}(y_d,y_s,y_b)$. Using a spurion analysis one can estimate
\beq
	c^{(3,1)}_{Q_{ij} L_{22}} \sim \left( \mathbf{1} + \alpha Y_u Y_u^\dagger + \beta Y_d Y_d^\dagger \right)_{i j}~,
\eeq
where $\alpha,\beta\sim \mathcal{O}(1)$, which implies the following structure:
\beq\begin{split}
	&C_{u\mu} = C_{c\mu} = C_{t\mu} \equiv C_{U\mu}~, \\
	&C_{d\mu} = C_{s\mu} = C_{b\mu} \equiv C_{D\mu}~,
	\label{eq:MFVstructure}
\end{split}\eeq
while flavour-violating terms are expected to be CKM suppressed, for example $|C_{bs\mu}| \sim |V_{tb} V_{ts}^* y_t^2 C_{D\mu}|$.
In this case the contribution to rare $B$ meson decays has a $V_{t s}$ suppression, while the dilepton signal at high-$p_T$ receives an universal contribution dominated by the valence quarks in the proton.
The flavour fit in Eq.~\eqref{eq:DC9} combined with this flavour structure would imply a value of $|C_{D\mu}| \sim 1.4 \times 10^{-3}$ which, as can be seen from the limits in Fig.~\ref{fig:MFV}, is already probed by the ATLAS dimuon search~\cite{ATLAS:2017wce} depending on the origin of the operator (i.e. from the SU(2) singlet or triplet structure) and will definitely be investigated at high luminosity.\footnote{It should also be noted that the triplet combination is bounded from the semileptonic hadron decays (CKM unitarity test) $C_{U\mu}-C_{D\mu}=(0.46\pm0.52)\times10^{-3}$~\cite{Gonzalez-Alonso:2016sip}, in the absence of other competing contributions.}
Allowing for more freedom and setting $C_{bs\mu} \equiv \lambda_{bs} C_{D\mu}$, we show in the top (central) panel of Fig.~\ref{fig:curtain} the 95\% CL limit in the $C_{D\mu}$-$|\lambda_{bs}|$ plane, where $C_{U\mu}$ is {related to $C_{D\mu}$ by assuming the triplet (singlet) structure}.
As discussed before, a direct upper limit on $\lambda_{bs}$, via $b-s$ fusion, can be derived only for very large values.
On the other hand, requiring $C_{bs\mu}$ to fit the $B$ decay anomalies already probes interesting regions in parameter space, excluding the MFV scenario ($\lambda_{bs} = V_{ts}$) for both singlet and triplet cases.

{\bf 2) $U(2)_Q$ flavour symmetry}\\
This symmetry distinguishes light left-handed quarks (doublets) from third generation left-handed quarks (singlets). The leading symmetry-breaking spurion is a doublet, whose flavour structure is unambiguously related to the CKM matrix~\cite{Barbieri:2011ci}.
In this case, in general the leading terms would involve the third generation quarks, as well as diagonal couplings in the first two generations. The relevant parameters for the dimuon production would then be
\beq\begin{split}
	C_{u\mu} = C_{c\mu} \equiv C_{U\mu}~, &\qquad
	C_{d\mu} = C_{s\mu} \equiv C_{D\mu}~, \\
	C_{b\mu}, &\qquad
	C_{bs\mu} \equiv \lambda_{bs} C_{b\mu}~,
	\label{eq:U2structure}
\end{split}\eeq
where the flavour violating coupling is expected to be $|\lambda_{bs}| \sim |V_{ts}|$.
As already done in the MFV case, in the following we leave $\lambda_{bs}$ free to vary and perform a four-parameter fit to the dimuon spectrum.
The resulting limits on $C_{U\mu}$ and $C_{D\mu}$ are very similar to those obtained in the MFV scenario (see Fig.~\ref{fig:MFV}) {and are required to be much smaller than the allowed range for $C_{b\mu}$.}

In the lower panel of Fig.~\ref{fig:curtain} we show the present and projected limits in the $C_{b\mu}$-$\lambda_{bs}$ plane (here we set $C_{D\mu} = C_{U\mu} = 0$, after checking that no large correlation with them is present). As for the MFV case, the fit of the flavour anomalies in Eq.~\eqref{eq:DC9}, combined with the upper limit on $|C_{b\mu}|$, provides a lower bound on $|\lambda_{bs}|$.
In this case, while at present this limit is much lower than the natural value predicted from $U(2)$ symmetry, $\lambda_{bs} \sim V_{ts}$, with high luminosity an interesting region will be probed. For example, in the $U(2)$ flavour models of Ref.~\cite{Greljo:2015mma,Buttazzo:2016kid,Bordone:2017anc,Barbieri:2015yvd} a small value of $\lambda_{bs}$ is necessary in order to pass the bounds from $B-\bar{B}$ mixing.

{\bf 3) Single-operator benchmarks}:

It is illustrative to show the limits on $\lambda^q_{bs}$ when only one {flavour-diagonal coefficient $C_{q\mu}$ is non-vanishing, while fitting at the same time $\Delta C^\mu_9$ in Eq.~\eqref{eq:DC9}. The expected $2\sigma$  limits with $36.1$~fb$^{-1}$ ($3000$~fb$^{-1}$) are:}
\beq
\begin{split}
\lambda^u_{bs} &> 0.072~(0.77), \; \; \lambda^u_{bs} < -0.097~(-0.76)~,\\
\lambda^d_{bs} &> 0.049~(0.36), \; \; \lambda^d_{bs} < -0.032~(-0.34)~,\\
\lambda^s_{bs} &> 0.007~(0.04), \; \; \lambda^s_{bs} < -0.004~(-0.03)~,\\
\lambda^c_{bs} &> 0.003~(0.02), \; \; \lambda^c_{bs} < -0.004~(-0.02)~,\\
\lambda^b_{bs} &> 0.002~(0.01), \; \; \lambda^b_{bs} < -0.002~(-0.006)~.
\end{split}
\eeq

\subsection{Model examples}

Let us briefly speculate about the UV scenarios capable of explaining the observed pattern of deviations in the rare $B$ meson decays.
For our EFT approach to be valid, we focus on models with new resonances beyond the kinematical reach for threshold production at the LHC. 
In such models, the effective operators in Eq.~\eqref{eq:opSMEFT} are presumably generated at the tree level.\footnote{Note that including a loop suppression factor of $\sim \frac{1}{16 \pi^2}$, the fit of the flavour anomalies in Eq.~\eqref{eq:DC9} points to a scale $\Lambda \approx 2.6^{+0.2}_{-0.3}$~TeV (see for example models proposed in Refs.~\cite{Kamenik:2017tnu,Arnan:2016cpy,Becirevic:2017jtw}).}
We focus here on the single mediator models in which the required effect is obtained by integrating out a single resonance. These include either an extra $Z'$ bosons~\cite{Greljo:2015mma,Buttazzo:2016kid,Gauld:2013qja,Buras:2013dea,Altmannshofer:2014cfa,Crivellin:2015mga,Crivellin:2015lwa,Celis:2015ara,Falkowski:2015zwa,Boucenna:2016wpr,Crivellin:2016ejn,Allanach:2015gkd,Chiang:2016qov,Alonso:2017bff} or a leptoquark~\cite{Hiller:2014ula,Varzielas:2015iva,Fajfer:2015ycq,Gripaios:2014tna,Barbieri:2015yvd,Barbieri:2016las,Alonso:2015sja,Pas:2015hca,Bauer:2015knc,Hiller:2017bzc} (for a recent review on leptoquarks see~\cite{Dorsner:2016wpm}). 

We note that a full set of single mediator models with tree-level matching to the vector triplet ($c^{(3)}_{Q_{ij} L_{kl}}$) or singlet ($c^{(1)}_{Q_{ij} L_{kl}}$) operators, consists of: color-singlet vectors $Z'_\mu \sim ({\bf 1},{\bf 1},0)$ and $W'_\mu \sim ({\bf 1},{\bf 3},0)$, color-triplet scalar $S_3 \sim (\bar {\bf 3},{\bf 3},1/3)$, and vectors $U_1^\mu \sim ({\bf 3},{\bf 1},2/3)$, $U_3^\mu \sim ({\bf 3},{\bf 3},2/3)$, in the notation of Ref.~\cite{Dorsner:2016wpm}. The quantum numbers in brackets indicate color, weak, and hypercharge representations, respectively.

\begin{figure}[t]
\centering
\includegraphics[width=0.9\hsize]{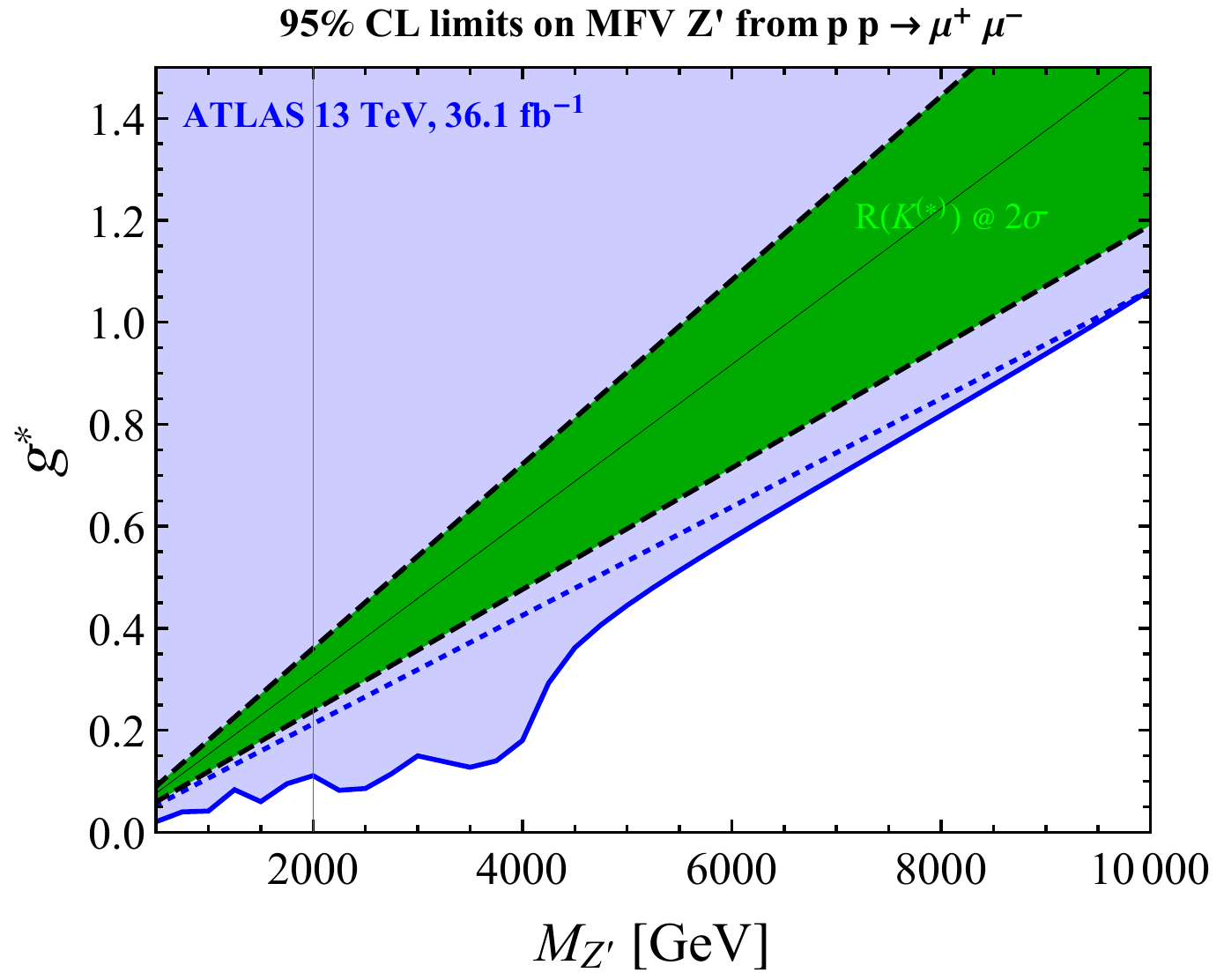}
\caption{Limits on the $Z'$ MFV model from $p p \to \mu^+ \mu^-$. See text for details.}\label{fig:Zprime}
\end{figure}

{\bf $Z'$ and $W'$ models}: A color-singlet vector resonance gives rise to an $s$-channel {resonant contribution to} the dilepton invariant mass distributions if $M_{Z'}$ is kinematically accessible. Otherwise, the deviation in the tails is {described well by} the dimension-six operators in Eq.~\eqref{eq:opSMEFT} with $\Lambda = M_V$ and
\beq
	c^{(3)}_{Q_{ij} L_{kl}} = - g_Q^{(3), i j} g_L^{(3), k l}~, \quad
	c^{(1)}_{Q_{ij} L_{kl}} = - g_Q^{(1), i j} g_L^{(1), k l}~,
\eeq
obtained after integrating out the heavy vectors with interactions $\mathcal{L} \supset Z'_\mu J_\mu + W'^{a}_\mu J^{a}_\mu$, where
\beq
\begin{split}
J_\mu &= g_Q^{(1),i j} (\bar Q_i \gamma_\mu Q_j) +g_L^{(1),k l}(\bar L_k \gamma^\mu L_l)~,\\
J^a_\mu &= g_Q^{(3),i j} (\bar Q_i \gamma_\mu \sigma^a Q_j) +g_L^{(3),k l}(\bar L_k \gamma^\mu \sigma^a L_l)~.
\end{split}
\label{eq:ZprimeCurr}
\eeq
A quark flavour-violating $g_Q^{(x),2 3}$ coupling and $g_L^{(x),22}$ are required to explain the flavour anomalies, while the limits from $p p \to \mu^+ \mu^-$ reported in Table~\ref{tab:Fit1}, can easily be translated to the flavour-diagonal couplings and mass combinations.

For example, assuming a singlet $Z'$ with $g_Q^{1,i j} = g_L^{1,i j} = \delta^{ij} g_*$ and MFV structure ($g_Q^{(1),2 3} = V_{ts} g_*$) we derive limits on $g_*$ as a function of the mass $M_{Z'}$, both fitting {the data} directly in the full model,\footnote{The $Z'$ decay width is determined by decays into the SM fermions $u,d,s,c,b,t,\mu,\nu_{\mu}$ via Eq.~\eqref{eq:ZprimeCurr}, i.e. $\Gamma_{Z'} / M_{Z'} = 5 g_*^2 / (6 \pi)$.} and in the EFT approach. The results are shown in Fig.~\ref{fig:Zprime}. The limits in the full model are shown with solid-blue while those in the EFT are shown with dashed-blue.
We see that for a mass $M_{Z'} \gtrsim 4-5$ TeV the limits in the two approaches agree well, while for the lower masses the EFT still provides conservative bounds.\footnote{See Ref.~\cite{Farina:2016rws} for a more detailed discussion on the EFT validity in high-$p_T$ dilepton tails.}
On top of this, we show with green lines the best fit and 2$\sigma$ interval which reproduce the $b \to s \mu\mu$ flavour anomalies, showing how LHC dimuon searches already exclude such a scenario independently of the $Z'$ mass.

Related to the above analysis, let us comment on the model recently proposed in Ref.~\cite{Alonso:2017bff}. An anomaly-free horizontal gauge symmetry is introduced, with a corresponding gauge field ($Z'_h$) having MFV-like couplings in the quark sector. Fig.~1 of Ref.~\cite{Alonso:2017bff} shows the preferred region from $\Delta C_9^\mu$ in the mass versus coupling plane, as well as the constraint from the $Z'$ resonance search (from the same experimental analysis used here~\cite{ATLAS:2017wce}). While the limits from the resonance search are effective up to $\sim 4$~TeV, we note that the limits from the tails go even beyond and already probe the interesting parameter region as shown in our Fig.~\ref{fig:curtain}. Note that this statement is independent of the $Z'$ mass (as long as the EFT is valid). 

{\bf Leptoquark models:}
A color-triplet resonance in the $t$-channel gives rise to $p p \to \ell^+ \ell^-$ at the LHC~\cite{Davidson:2014lsa,Bessaa:2014jya}. The relevant interaction Lagrangian for explaining $B$ decay anomalies is,
\beq
\begin{aligned}
\mathcal{L} &\supset y_{3 i j}^{L L} \bar{Q}^{c, i}_L i \sigma^2 \sigma^a L^{j}_L S^a_3 + x_{3 i j}^{LL} \bar Q_L^i \gamma^\mu \sigma^a L_L^j U_{3,\mu}^a \\
&+x_{1 i j}^{L L} \bar Q_L^i \gamma^\mu L_L^j U_{1,\mu}+ \rm{h.c.}~,
\end{aligned}
\eeq
and the matching to the EFT is provided in Table~4 of Ref.~\cite{Dorsner:2016wpm}.
The constraints from Table~\ref{tab:Fit1} apply again in a straightforward way. The validity of the expansion has been studied in details in Refs.~\cite{Davidson:2014lsa,Bessaa:2014jya}. We would like to point out that similar limits would apply even for a relatively light LQ (in the $\sim$~TeV range). As an illustration, the fit to low-energy anomalies in the model of Ref.~\cite{Becirevic:2017jtw} (where the effect is loop-generated), requires large charm-muon-LQ coupling, leading to a potentially observable $c ~ \bar c \to \mu^+ \mu^-$ production at high-$p_T$. We also note that the single LQ production at the LHC can constrain similar couplings~\cite{Dorsner:2014axa}.

\section{Conclusions}
\label{sec:conclusions}

In this work we discuss the contribution from flavour non-universal new physics to the high-$p_T$ dilepton tails in $p p \to \ell^+ \ell^-$, where $\ell=e,\mu$. In particular, we set the best up-to-date limits on all 36 four-fermion operators in the SMEFT which contribute to these processes by recasting the recent 13~TeV ATLAS analysis with 36.1~fb$^{-1}$ of data, as well as estimate the final sensitivity for the high-luminosity phase at the LHC.

Recent results in rare semileptonic $B$ meson decays show some intriguing hints for possible violation of lepton-flavour universality. It is particularly interesting to notice that all the different anomalies can be coherently described by a new physics contribution to the left-handed $b_L \to s_L \mu_L^+ \mu_L^-$ contact interaction. In most flavour models, the flavour-changing interactions are related (and usually suppressed with respect) to the flavour diagonal ones. These, in turn, are probed via the high-$p_T$ dimuon tail, allowing us to set limits which are already probing interesting regions of parameter space of some models.

In particular, our limits exclude, or put in strong tension, scenarios which aim to describe the flavour anomalies using MFV structure that directly relates the $bs\mu\mu$ contact interaction to the ones involving first generation quarks, tightly constrained from $p p \to \mu^+ \mu^-$.
On the other hand, scenarios with $U(2)_Q$ flavour symmetry predominantly coupled to the third generation quarks lead to milder constraints. We also briefly discuss a few explicit examples with heavy mediator states (colourless vectors and leptoquarks), and show a comparison of the limits obtained in the EFT with those obtained directly in the model.

If these flavour anomalies will be confirmed with more data, correlated signals at high-$p_T$ processes at LHC will be crucial in order to decipher the responsible dynamics. We show that the high energy dilepton tails can provide very valuable information in this direction.


\begin{acknowledgements}
We would like to thank Mart\'in Gonz\'alez-Alonso and Gino Isidori for useful discussions. This work is supported in part by the Swiss National Science Foundation (SNF) under contract 200021-159720.
\end{acknowledgements}

\begin{appendix}
\section{dilepton cross section}
\label{app:sigma}

The unpolarized partonic differential cross section following from Eq.~\eqref{eq:1} is given by
\beq
\begin{split}
\frac{d \hat \sigma}{d t} &= \frac{1}{48 \pi s^2}~ u^2 \left(|F_{q_L \ell_L}(s)|^2+|F_{q_R \ell_R}(s)|^2\right)\\
 & + \frac{1}{48 \pi s^2}~ t^2 \left(|F_{q_L \ell_R}(s)|^2+|F_{q_R \ell_L}(s)|^2\right)~,
 \end{split}
\eeq
where $s$, $t$, and $u$ are the Mandelstam variables. 
The total partonic cross section is
{\footnotesize
\beq
\hat \sigma = \frac{s}{144 \pi} \left(|F_{q_L \ell_L}(s)|^2+|F_{q_R \ell_R}(s)|^2+|F_{q_L \ell_R}(s)|^2+|F_{q_R \ell_L}(s)|^2 \right)~,
\eeq}
\!\!\!while the hadronic cross section is obtained after convoluting the partonic one with the corresponding parton luminosity functions
\beq
\mathcal{L}_{q \bar q}(\tau,\mu_F) = \int_{\tau}^1 \frac{d x}{x}~f_q (x,\mu_F) f_{\bar q} (\tau/x,\mu_F) ~.
\eeq
In particular, the cross section in the dilepton invariant mass bin $\left[\tau_{\rm{min}}^{\rm{bin}}, \tau_{\rm{max}}^{\rm{bin}}\right]$ is given by
\beq
\sigma^{{\rm bin}}(p~ p \to \ell^+ \ell^-) =  \sum_q ~ \int_{\tau_{\rm{min}}^{\rm{bin}}}^{\tau_{\rm{max}}^{\rm{bin}}} d \tau ~ 2 \mathcal{L}_{q \bar q}(\tau,\mu_F)~\hat \sigma(\tau s_0)~.
\label{eq:tot-cross}
\eeq

\section{Operator limits}
\label{app:Limits}
In Table~\ref{tab:Fit1} we show the present 2$\sigma$ limits on the 36 independent four-fermion operators contributing to $p p \to \ell^+ \ell^-$ from the 13~TeV ATLAS analysis~\cite{ATLAS:2017wce} with 36.1~fb$^{-1}$ of data, as well as projections for 3000~fb$^{-1}$, where only one operator is turned on at a time.
The notation used is as in Eq.~\eqref{eq:opSMEFT} but the cutoff dependence has been reabsorbed as $C_x \equiv \frac{v^2}{\Lambda^2} c_x$. 
In the case of operators involving $b_L$ quark, instead, we keep only the combination of triplet and singlet aligned with it, since the top quark does not enter in this observable.
In the Gaussian approximation we derived the correlation matrix in the 36 coefficients and checked that the only non-negligible correlation is the one among the triplet and singlet $(\bar{L}L)(\bar{L}L)$ operators with same fermion content. This correlation is shown explicitly in the 2d fit of Fig.~\ref{fig:MFV}.

\begin{table}[t]
\renewcommand{\arraystretch}{1.35}
\begin{tabular}{c|c|c}
$C_i $ & ATLAS 36.1 fb$^{-1}$ & 3000 fb$^{-1}$ \\\hline
  $C^{(1)}_{Q^1 L^2}$ &~ [-5.73, 14.2] $\times 10^{-4}$ &~ [-1.30, 1.51] $\times 10^{-4}$ \\
  $C^{(3)}_{Q^1 L^2}$ &~ [-7.11, 2.84] $\times 10^{-4}$ &~ [-5.25, 5.25] $\times 10^{-5}$ \\
  $C_{u_R L^2}$ &~ [-0.84, 1.61] $\times 10^{-3}$ &~ [-2.00, 2.66] $\times 10^{-4}$ \\
  $C_{u_R \mu_R}$ &~ [-0.52, 1.36] $\times 10^{-3}$ &~ [-1.04, 1.08] $\times 10^{-4}$ \\
  $C_{Q^1 \mu_R}$ &~ [-0.82, 1.27] $\times 10^{-3}$ &~ [-2.25, 4.10] $\times 10^{-4}$ \\
  $C_{d_R L^2}$ &~ [-2.13, 1.61] $\times 10^{-3}$ &~ [-8.98, 5.11] $\times 10^{-4}$ \\
  $C_{d_R \mu_R}$ &~ [-2.31, 1.34] $\times 10^{-3}$ &~ [-4.89, 3.33] $\times 10^{-4}$ \\
  $C^{(1)}_{Q^2 L^2}$ &~ [-8.84, 7.35] $\times 10^{-3}$ &~ [-3.83, 2.39] $\times 10^{-3}$ \\
  $C^{(3)}_{Q^2 L^2}$ &~ [-9.75, 5.56] $\times 10^{-3}$ &~ [-1.43, 1.15] $\times 10^{-3}$ \\
  $C_{Q^2 \mu_R}$ &~ [-7.53, 8.67] $\times 10^{-3}$ &~ [-2.58, 3.73] $\times 10^{-3}$ \\
  $C_{s_R L^2}$ &~ [-1.04 , 0.93] $\times 10^{-2}$ &~ [-4.42, 3.33] $\times 10^{-3}$ \\
  $C_{s_R \mu_R}$ &~ [-1.09 , 0.87] $\times 10^{-2}$ &~ [-4.67, 2.73] $\times 10^{-3}$ \\
  $C_{c_R L^2}$ &~ [-1.33, 1.52] $\times 10^{-2}$ &~ [-4.58, 6.54] $\times 10^{-3}$ \\
  $C_{c_R \mu_R}$ &~ [-1.21, 1.62] $\times 10^{-2}$ &~ [-3.48, 6.32] $\times 10^{-3}$ \\
  $C_{b_L L^2}$ &~ [-2.61, 2.07] $\times 10^{-2}$ &~ [-11.1, 6.33] $\times 10^{-3}$ \\
  $C_{b_L \mu_R}$ &~ [-2.28, 2.42] $\times 10^{-2}$ &~ [-8.53, 10.0] $\times 10^{-3}$ \\
  $C_{b_R L^2}$ &~ [-2.41, 2.29] $\times 10^{-2}$ &~ [-9.90, 8.68] $\times 10^{-3}$ \\
  $C_{b_R \mu_R}$ &~ [-2.47, 2.23] $\times 10^{-2}$ &~ [-10.5, 7.97] $\times 10^{-3}$
 \end{tabular}
 \begin{tabular}{c|c|c}
$C_i $ & ATLAS 36.1 fb$^{-1}$ & 3000 fb$^{-1}$ \\\hline
  $C^{(1)}_{Q^1 L^1}$ &~ [-0.0, 1.75] $\times 10^{-3}$ &~ [-1.01, 1.13] $\times 10^{-4}$ \\
  $C^{(3)}_{Q^1 L^1}$ &~ [-8.92, -0.54] $\times 10^{-4}$ &~ [-3.99, 3.93] $\times 10^{-5}$ \\
  $C_{u_R L^1}$ &~ [-0.19, 1.92] $\times 10^{-3}$ &~ [-1.56, 1.92] $\times 10^{-4}$ \\
  $C_{u_R e_R}$ &~ [0.15, 2.06] $\times 10^{-3}$ &~ [-7.89, 8.23] $\times 10^{-5}$ \\
  $C_{Q^1 e_R}$ &~ [-0.40, 1.37] $\times 10^{-3}$ &~ [-1.8, 2.85] $\times 10^{-4}$ \\
  $C_{d_R L^1}$ &~ [-2.1, 1.04] $\times 10^{-3}$ &~ [-7.59, 4.23] $\times 10^{-4}$ \\
  $C_{d_R e_R}$ &~ [-2.55, 0.46] $\times 10^{-3}$ &~ [-3.37, 2.59] $\times 10^{-4}$ \\
  $C^{(1)}_{Q^2 L^1}$ &~ [-6.62, 4.36] $\times 10^{-3}$ &~ [-3.31, 1.92] $\times 10^{-3}$ \\
  $C^{(3)}_{Q^2 L^1}$ &~ [-8.24, 2.05] $\times 10^{-3}$ &~ [-8.87, 7.90] $\times 10^{-4}$ \\
  $C_{Q^2 e_R}$ &~ [-4.67, 6.34] $\times 10^{-3}$ &~ [-2.11, 3.30] $\times 10^{-3}$ \\
  $C_{s_R L^1}$ &~ [-7.4 , 5.9] $\times 10^{-3}$ &~ [-3.96, 2.8] $\times 10^{-3}$ \\
  $C_{s_R e_R}$ &~ [-8.17, 5.06] $\times 10^{-3}$ &~ [-3.82, 2.13] $\times 10^{-3}$ \\
  $C_{c_R L^1}$ &~ [-0.83, 1.13] $\times 10^{-2}$ &~ [-3.74, 5.77] $\times 10^{-3}$ \\
  $C_{c_R e_R}$ &~ [-0.67, 1.27] $\times 10^{-2}$ &~ [-2.59, 4.17] $\times 10^{-3}$ \\
  $C_{b_L L^1}$ &~ [-1.93, 1.19] $\times 10^{-2}$ &~ [-8.62, 4.82] $\times 10^{-3}$ \\
  $C_{b_L e_R}$ &~ [-1.47, 1.67] $\times 10^{-2}$ &~ [-7.29, 8.99] $\times 10^{-3}$ \\
  $C_{b_R L^1}$ &~ [-1.65, 1.49] $\times 10^{-2}$ &~ [-8.86, 7.48] $\times 10^{-3}$ \\
  $C_{b_R e_R}$ &~ [-1.73, 1.40] $\times 10^{-2}$ &~ [-9.38, 6.63] $\times 10^{-3}$
 \end{tabular}
 \caption{\label{tab:Fit1} One-parameter 2$\sigma$ limits from $p p \to \mu^+ \mu^-, e^+ e^-$.}
\end{table}

\end{appendix}


\begin{thebibliography}{99}        
\addcontentsline{toc}{section}{References}

\bibitem{Isidori:2010kg}
  G.~Isidori, Y.~Nir and G.~Perez,
  Ann.\ Rev.\ Nucl.\ Part.\ Sci.\  {\bf 60} (2010) 355
  [\arXhref{arXiv:1002.0900} [hep-ph]].
  
\bibitem{Aaij:2014ora}
  R.~Aaij {\it et al.} [LHCb Collaboration],
  Phys.\ Rev.\ Lett.\  {\bf 113} (2014) 151601
  [\arXhref{arXiv:1406.6482} [hep-ex]].
  
\bibitem{Aaij:2013qta}
  R.~Aaij {\it et al.} [LHCb Collaboration],
  Phys.\ Rev.\ Lett.\  {\bf 111} (2013) 191801
  [\arXhref{arXiv:1308.1707} [hep-ex]].
  
\bibitem{Aaij:2015oid}
  R.~Aaij {\it et al.} [LHCb Collaboration],
  JHEP {\bf 1602} (2016) 104
  [\arXhref{arXiv:1512.04442} [hep-ex]].
  
\bibitem{BifaniTALK}  
S. Bifani, in Seminar at CERN, April 18th (2017).

\bibitem{Cirigliano:2012ab}
  V.~Cirigliano, M.~Gonzalez-Alonso and M.~L.~Graesser,
  JHEP {\bf 1302} (2013) 046
  [\arXhref{arXiv:1210.4553} [hep-ph]].

\bibitem{Gonzalez-Alonso:2016sip}
  M.~González-Alonso and J.~Martin Camalich,
  \arXhref{arXiv:1606.06037} [hep-ph].

\bibitem{deBlas:2013qqa}
  J.~de Blas, M.~Chala and J.~Santiago,
  Phys.\ Rev.\ D {\bf 88} (2013) 095011
  [\arXhref{arXiv:1307.5068} [hep-ph]].

\bibitem{Farina:2016rws}
  M.~Farina, G.~Panico, D.~Pappadopulo, J.~T.~Ruderman, R.~Torre and A.~Wulzer,
  \arXhref{arXiv:1609.08157} [hep-ph].

\bibitem{Faroughy:2016osc}
  D.~A.~Faroughy, A.~Greljo and J.~F.~Kamenik,
  Phys.\ Lett.\ B {\bf 764} (2017) 126
  [\arXhref{arXiv:1609.07138} [hep-ph]].
 
\bibitem{ATLAS:2017wce}
  The ATLAS collaboration [ATLAS Collaboration],
  ATLAS-CONF-2017-027.
  
  
\bibitem{Grzadkowski:2010es}
  B.~Grzadkowski, M.~Iskrzynski, M.~Misiak and J.~Rosiek,
  JHEP {\bf 1010} (2010) 085
  [\arXhref{arXiv:1008.4884} [hep-ph]].
 
\bibitem{Efrati:2015eaa}
  A.~Efrati, A.~Falkowski and Y.~Soreq,
  JHEP {\bf 1507} (2015) 018
  [\arXhref{arXiv:1503.07872} [hep-ph]].
 
 
\bibitem{Cowan:2010js}
  G.~Cowan, K.~Cranmer, E.~Gross and O.~Vitells,
  Eur.\ Phys.\ J.\ C {\bf 71} (2011) 1554
   Erratum: [Eur.\ Phys.\ J.\ C {\bf 73} (2013) 2501]
  [\arXhref{arXiv:1007.1727} [physics.data-an]].
 
\bibitem{Harland-Lang:2014zoa}
  L.~A.~Harland-Lang, A.~D.~Martin, P.~Motylinski and R.~S.~Thorne,
  Eur.\ Phys.\ J.\ C {\bf 75} (2015) no.5,  204
  [\arXhref{arXiv:1412.3989} [hep-ph]].

\bibitem{Alloul:2013bka}
  A.~Alloul, N.~D.~Christensen, C.~Degrande, C.~Duhr and B.~Fuks,
  Comput.\ Phys.\ Commun.\  {\bf 185} (2014) 2250
  [\arXhref{arXiv:1310.1921} [hep-ph]].

\bibitem{Alwall:2014hca}
  J.~Alwall {\it et al.},
  JHEP {\bf 1407} (2014) 079
  [\arXhref{arXiv:1405.0301} [hep-ph]].
  
\bibitem{Capdevila:2017bsm}
  B.~Capdevila, A.~Crivellin, S.~Descotes-Genon, J.~Matias and J.~Virto,
  \arXhref{arXiv:1704.05340} [hep-ph].
 
\bibitem{Altmannshofer:2017yso}
  W.~Altmannshofer, P.~Stangl and D.~M.~Straub,
  \arXhref{arXiv:1704.05435} [hep-ph].

\bibitem{Geng:2017svp}
  L.~S.~Geng, B.~Grinstein, S.~Jager, J.~Martin Camalich, X.~L.~Ren and R.~X.~Shi,
  \arXhref{arXiv:1704.05446} [hep-ph].

\bibitem{DescotesGenon:2012zf}
  S.~Descotes-Genon, J.~Matias, M.~Ramon and J.~Virto,
  JHEP {\bf 1301} (2013) 048
  [\arXhref{arXiv:1207.2753} [hep-ph]].

\bibitem{Hiller:2003js}
  G.~Hiller and F.~Kruger,
  Phys.\ Rev.\ D {\bf 69} (2004) 074020
  [\arXhref{hep-ph/0310219}].
  
\bibitem{Bobeth:2007dw}
  C.~Bobeth, G.~Hiller and G.~Piranishvili,
  JHEP {\bf 0712} (2007) 040
  [\arXhref{arXiv:0709.4174} [hep-ph]].

\bibitem{Bordone:2016gaq}
  M.~Bordone, G.~Isidori and A.~Pattori,
  Eur.\ Phys.\ J.\ C {\bf 76} (2016) no.8,  440
  [\arXhref{arXiv:1605.07633} [hep-ph]].

\bibitem{Celis:2017doq} 
  A.~Celis, J.~Fuentes-Martin, A.~Vicente and J.~Virto,
  \arXhref{arXiv:1704.05672} [hep-ph].

\bibitem{Bhattacharya:2014wla}
  B.~Bhattacharya, A.~Datta, D.~London and S.~Shivashankara,
  Phys.\ Lett.\ B {\bf 742} (2015) 370
  [\arXhref{arXiv:1412.7164} [hep-ph]].

\bibitem{Alonso:2015sja}
  R.~Alonso, B.~Grinstein and J.~Martin Camalich,
  JHEP {\bf 1510} (2015) 184
  [\arXhref{arXiv:1505.05164} [hep-ph]].

\bibitem{Greljo:2015mma}
  A.~Greljo, G.~Isidori and D.~Marzocca,
  JHEP {\bf 1507} (2015) 142
  [\arXhref{arXiv:1506.01705} [hep-ph]].

\bibitem{Olive:2016xmw}
  C.~Patrignani {\it et al.} [Particle Data Group],
  Chin.\ Phys.\ C {\bf 40} (2016) no.10,  100001.

\bibitem{DAmbrosio:2002vsn}
  G.~D'Ambrosio, G.~F.~Giudice, G.~Isidori and A.~Strumia,
  Nucl.\ Phys.\ B {\bf 645} (2002) 155
  [\arXhref{hep-ph/0207036}].

\bibitem{Barbieri:2011ci}
  R.~Barbieri, G.~Isidori, J.~Jones-Perez, P.~Lodone and D.~M.~Straub,
  Eur.\ Phys.\ J.\ C {\bf 71} (2011) 1725
  [\arXhref{arXiv:1105.2296} [hep-ph]].

\bibitem{Buttazzo:2016kid}
  D.~Buttazzo, A.~Greljo, G.~Isidori and D.~Marzocca,
  JHEP {\bf 1608} (2016) 035
  [\arXhref{arXiv:1604.03940} [hep-ph]].


\bibitem{Bordone:2017anc}
  M.~Bordone, G.~Isidori and S.~Trifinopoulos,
\arXhref{arXiv:1702.07238} [hep-ph].


 
\bibitem{Kamenik:2017tnu}
  J.~F.~Kamenik, Y.~Soreq and J.~Zupan,
  \arXhref{arXiv:1704.06005} [hep-ph].
   
\bibitem{Arnan:2016cpy}
  P.~Arnan, L.~Hofer, F.~Mescia and A.~Crivellin,
  JHEP {\bf 1704} (2017) 043
  doi:10.1007/JHEP04(2017)043
  [  \arXhref{arXiv:1608.07832} [hep-ph]].
   
\bibitem{Becirevic:2017jtw}
  D.~Be\v cirevi\' c and O.~Sumensari,
  \arXhref{arXiv:1704.05835} [hep-ph].

  
\bibitem{Gauld:2013qja}
  R.~Gauld, F.~Goertz and U.~Haisch,
  JHEP {\bf 1401} (2014) 069
  [\arXhref{arXiv:1310.1082} [hep-ph]].
  
\bibitem{Buras:2013dea}
  A.~J.~Buras, F.~De Fazio and J.~Girrbach,
  JHEP {\bf 1402} (2014) 112
  [\arXhref{arXiv:1311.6729} [hep-ph]].
 
\bibitem{Altmannshofer:2014cfa} 
  W.~Altmannshofer, S.~Gori, M.~Pospelov and I.~Yavin,
  Phys.\ Rev.\ D {\bf 89}, 095033 (2014)
  [\arXhref{arXiv:1403.1269} [hep-ph]].
  
\bibitem{Crivellin:2015mga}
  A.~Crivellin, G.~D'Ambrosio and J.~Heeck,
  Phys.\ Rev.\ Lett.\  {\bf 114} (2015) 151801
  [\arXhref{arXiv:1501.00993} [hep-ph]].
  
\bibitem{Crivellin:2015lwa}
  A.~Crivellin, G.~D'Ambrosio and J.~Heeck,
  Phys.\ Rev.\ D {\bf 91} (2015) no.7,  075006
  [\arXhref{arXiv:1503.03477} [hep-ph]].
 
\bibitem{Celis:2015ara}
  A.~Celis, J.~Fuentes-Martin, M.~Jung and H.~Serodio,
  Phys.\ Rev.\ D {\bf 92} (2015) no.1,  015007
  [\arXhref{arXiv:1505.03079} [hep-ph]].
 
\bibitem{Falkowski:2015zwa}
  A.~Falkowski, M.~Nardecchia and R.~Ziegler,
  JHEP {\bf 1511} (2015) 173
  [\arXhref{arXiv:1509.01249} [hep-ph]].
 
\bibitem{Boucenna:2016wpr}
  S.~M.~Boucenna, A.~Celis, J.~Fuentes-Martin, A.~Vicente and J.~Virto,
  Phys.\ Lett.\ B {\bf 760} (2016) 214
  [\arXhref{arXiv:1604.03088} [hep-ph]].
 
\bibitem{Crivellin:2016ejn}
  A.~Crivellin, J.~Fuentes-Martin, A.~Greljo and G.~Isidori,
  Phys.\ Lett.\ B {\bf 766} (2017) 77
  [\arXhref{arXiv:1611.02703} [hep-ph]].
 
\bibitem{Allanach:2015gkd}
  B.~Allanach, F.~S.~Queiroz, A.~Strumia and S.~Sun,
  Phys.\ Rev.\ D {\bf 93} (2016) no.5,  055045
  [\arXhref{arXiv:1511.07447} [hep-ph]].
 
\bibitem{Chiang:2016qov}
  C.~W.~Chiang, X.~G.~He and G.~Valencia,
  Phys.\ Rev.\ D {\bf 93} (2016) no.7,  074003
  [\arXhref{arXiv:1601.07328} [hep-ph]].
 
\bibitem{Alonso:2017bff}
  R.~Alonso, P.~Cox, C.~Han and T.~T.~Yanagida,
  \arXhref{arXiv:1704.08158} [hep-ph].
 
 
\bibitem{Hiller:2014ula}
  G.~Hiller and M.~Schmaltz,
  JHEP {\bf 1502} (2015) 055
  [\arXhref{arXiv:1411.4773} [hep-ph]].
 
\bibitem{Varzielas:2015iva}
  I.~de Medeiros Varzielas and G.~Hiller,
  JHEP {\bf 1506} (2015) 072
  [\arXhref{arXiv:1503.01084} [hep-ph]].

\bibitem{Fajfer:2015ycq}
  S.~Fajfer and N.~Ko¨nik,
  Phys.\ Lett.\ B {\bf 755} (2016) 270
  [\arXhref{arXiv:1511.06024} [hep-ph]].
 
\bibitem{Gripaios:2014tna}
  B.~Gripaios, M.~Nardecchia and S.~A.~Renner,
  JHEP {\bf 1505} (2015) 006
  [\arXhref{arXiv:1412.1791} [hep-ph]].

\bibitem{Barbieri:2015yvd}
  R.~Barbieri, G.~Isidori, A.~Pattori and F.~Senia,
  Eur.\ Phys.\ J.\ C {\bf 76} (2016) no.2,  67
  [\arXhref{arXiv:1512.01560} [hep-ph]].
 
\bibitem{Barbieri:2016las}
  R.~Barbieri, C.~W.~Murphy and F.~Senia,
  Eur.\ Phys.\ J.\ C {\bf 77} (2017) no.1,  8
  [\arXhref{arXiv:1611.04930} [hep-ph]].
 
\bibitem{Pas:2015hca}
  H.~Päs and E.~Schumacher,
  Phys.\ Rev.\ D {\bf 92} (2015) no.11,  114025
  [\arXhref{arXiv:1510.08757} [hep-ph]].

\bibitem{Bauer:2015knc}
  M.~Bauer and M.~Neubert,
  Phys.\ Rev.\ Lett.\  {\bf 116} (2016) no.14,  141802
  [\arXhref{arXiv:1511.01900} [hep-ph]].
 
\bibitem{Hiller:2017bzc}
  G.~Hiller and I.~Nisandzic,
  \arXhref{arXiv:1704.05444} [hep-ph].

\bibitem{Dorsner:2016wpm}
  I.~Dor\v sner, S.~Fajfer, A.~Greljo, J.~F.~Kamenik and N.~Ko\v snik,
  Phys.\ Rept.\  {\bf 641} (2016) 1
  [\arXhref{arXiv:1603.04993} [hep-ph]].
 
\bibitem{Davidson:2014lsa}
  S.~Davidson, S.~Descotes-Genon and P.~Verdier,
  Phys.\ Rev.\ D {\bf 91} (2015) no.5,  055031
  [\arXhref{arXiv:1410.4798} [hep-ph]].
   
\bibitem{Bessaa:2014jya}
  A.~Bessaa and S.~Davidson,
  Eur.\ Phys.\ J.\ C {\bf 75} (2015) no.2,  97
  [\arXhref{arXiv:1409.2372} [hep-ph]].
  
\bibitem{Dorsner:2014axa}
  I.~Dorsner, S.~Fajfer and A.~Greljo,
  JHEP {\bf 1410} (2014) 154
  [\arXhref{arXiv:1406.4831} [hep-ph]].
 
\end{thebibliography}
\end{document}